\documentclass{article}
\usepackage{graphicx}
\usepackage{cite}
\begin{document}

\title{\bf Harmonic oscillator with minimal length, minimal momentum, and maximal momentum uncertainties in SUSYQM framework}
\author{M. Asghari,$^a$ P. Pedram\thanks{Email: p.pedram@srbiau.ac.ir},$^a$ and K. Nozari$^b$
\\  {\small $^a$Department of Physics, Science and Research Branch, Islamic Azad University, Tehran, Iran}\\
    {\small $^b$Center for Excellence in Astronomy and Astrophysics (CEAAI-RIAAM)-Maragha, P.~O.~Box: 55134--441, Maragha, Iran}}

\maketitle \baselineskip 24pt

\begin{abstract}
We consider a Generalized Uncertainty Principle (GUP) framework
which predicts a maximal uncertainty in momentum and minimal
uncertainties both in position and momentum. We apply supersymmetric
quantum mechanics method and the shape invariance condition to
obtain the exact harmonic oscillator eigenvalues in this GUP
context. We find the supersymmetric partner Hamiltonians and show
that the harmonic oscillator belongs to a hierarchy of Hamiltonians
with a shift in momentum representation and different masses and
frequencies. We also study the effect of a uniform electric field on
the harmonic oscillator energy spectrum in this setup.

\vspace{.5cm} {\it Keywords}: Harmonic Oscillator; Generalized
Uncertainty Principle; Supersymmetric Quantum Mechanics.
\end{abstract}

\section{Introduction}
The assumption of the continuity of spacetime manifold may be broken
at high energy limit. In this limit the effects of gravity are so
important that would result in discreetness of the spacetime. In the
context of quantum gravity, the Heisenberg uncertainty principle
should be modified to the so-called Generalized Uncertainty
Principle (GUP). This generalization leads to a nonzero minimal
uncertainty in position measurements. In ordinary quantum mechanics
we can make $\Delta x$ arbitrarily small by letting $\Delta p$ grow
correspondingly, but in the GUP framework there exists a nonzero and
minimal position uncertainty.

Various candidates of quantum gravity such as the string theory,
loop quantum gravity, and quantum geometry
\cite{1,2,3,4,5,6,7,8,9,10,11,12,13,29,30,31,32,33,34,35,36,37,39}
all indicate the existence of a minimal measurable length of the
order of the Planck length $\ell
_{pl}=\sqrt{\frac{G\hbar}{c^{3}}}\approx 10^{-35}$m. Kempf et al.
proposed a GUP proposal that implies a minimal length and
constructed the Hilbert space representation of quantum mechanics in
this GUP framework \cite{12}. On the other hand, since the curvature
of spacetime is important at large distances and the notion of the
plane wave does not hold on the curved spacetime, the existence of a
nonzero minimal uncertainty in momentum is also inevitable
\cite{14}. The idea of a maximal observable momentum can also be
incorporated into this scenario based on the Doubly Special
Relativity (DSR) theories where the Planck energy (Planck momentum)
is considered as an additional invariant other than the velocity of
light \cite{18,19,20}. Ali, Das and Vagenas have recently proposed
the following commutation relation between position and momentum
operators that implies minimal length and maximal momentum
uncertainties:
\begin{equation}
[x,p]=i\hbar(1+2{\bar{\alpha}}^{2}p^{2}-\bar{\alpha}p),
\end{equation}
where its Hilbert space representation is also constructed in
Ref.~\cite{39}.

Supersymmetric Quantum Mechanics (SUSYQM) is an application of the
idea of supersymmetry to quantum mechanics. Supersymmetry proposes
that to each fermion there exists a boson and vice versa. So, we can
think of supersymmetry as proposing that a symmetry exists between
bosons and fermions, and that in nature, there are equal numbers of
fermion and boson states. SUSYQM involves pairs of Hamiltonians
which share a particular mathematical relationship, which are called
partner Hamiltonians. For every eigenstate of one Hamiltonian, its
partner Hamiltonian has a corresponding eigenstate with the same
energy (except possibly for zero energy eigenstates). This fact can
be exploited to deduce many properties of the eigenstate spectrum.
It is analogous to the original description of SUSY, which referred
to bosons and fermions. One can imagine a \emph{bosonic
Hamiltonian}, whose eigenstates are the various bosons of the
theory. The SUSY partner of this Hamiltonian would be
\emph{fermionic}, and its eigenstates would be the theory's
fermions. Each boson would have a fermionic partner of equal energy
but, in the relativistic world, energy and mass are interchangeable,
so we can just as easily say that the partner particles have equal
mass. For more details and also machinery of SUSYQM see
Refs.~\cite{s1,s2}.

The problem of the harmonic oscillator has attracted much attention
in various GUP frameworks. In Ref.~\cite{12} this problem is exactly
solved in the presence of a minimal length. Quesne and Tkachuk
studied this problem with nonzero uncertainties in both position and
momentum in the supersymmetric framework \cite{21}. We can use
SUSYQM to construct exact solutions of many quantum mechanical
problems. A certain class of exactly solvable potentials have a
property known as the shape invariance \cite{22,22-1,23}. The
potentials that their SUSY partner has the same spatial dependence
with possibly altered parameters are shape invariant potentials. If
a potential satisfies this condition, we can obtain the energy
eigenvalues and eigenfunctions without solving the differential
equation \cite{24,25}. The SUSYQM method plus the shape invariance
is related to the factorization method developed by Schr\"odinger,
Infeld and Hull \cite{26}.

In this Letter, we apply SUSYQM method and the notion of the shape
invariance on the eigenvalue problem of the harmonic oscillator in a
GUP framework that predicts maximal uncertainty in momentum and
minimal uncertainties in both position and momentum. In this case
$x$ and $p$ satisfy the following modified commutation relation
\begin{eqnarray}\label{xp1}
[x,p]=i\hbar(1+\bar{\rho} x^{2}+2{\bar{\alpha}}^{2}p^{2}-\bar{\alpha}p),
\end{eqnarray}
with $\bar{\rho}\geq0$, $\bar{\alpha}\geq0$, and
$\bar{\rho}\bar{\alpha}<\hbar^{-2}$. We obtain the supersymmetric
partner Hamiltonian of the harmonic oscillator and show that the
harmonic oscillator belongs to a hierarchy of Hamiltonians with a
shift in momentum representation and with different masses and
frequencies. Finally we study the effect of a uniform electric field
on the harmonic oscillator energy spectrum in this setup.

\section{One-dimensional harmonic oscillator in the GUP framework}
In this section, using the SUSYQM method, we obtain the harmonic
oscillator energy spectrum in the GUP framework which implies
maximal uncertainty in momentum and minimal uncertainties in both
position and momentum. First let us introduce dimensionless position
and momentum operators, $X=x/a$ and $P=pa/\hbar$, where
$a=\sqrt{\hbar/(m\omega)}$ is known as the oscillator's length.
These dimensionless operators, satisfy the following deformed
commutation relation
\begin{equation}\label{eq1xp}
[X,P]=i (1+\rho X^{2}+2\alpha^{2} P^{2}-\alpha P),
\end{equation}
where $\rho=\bar{\rho}\hbar/(m\omega)$ and
$\alpha=\bar{\alpha}\sqrt{m\hbar\omega}$, are the dimensionless
parameters.

%***************************
Notice that, minimal length, minimal momentum and maximal momentum
all stand for the uncertainties. These are not a test particle's
characteristics, but these are limitations on measurement of
position or momentum for a test particle. In other words, we do not
suppose a test particle that has minimal momentum and maximal
momentum at the same time. Minimal momentum and maximal momentum are
natural cutoffs on particle's momentum measurement. Indeed, a test
particle's momentum cannot be less than the minimal momentum and
cannot be larger than the maximal momentum. Note that it is
supposed nontrivially that minimal uncertainty in momentum
measurement leads to the existence of a minimal momentum for a test
particle. Also a test particle's momentum cannot be arbitrarily
imprecise and therefore there is an upper bound for momentum
fluctuations. We suppose, as a nontrivial assumption, that this
upper bound for momentum fluctuation defines a maximal measurable
momentum for a test particle. So, we can define a GUP that contains
all possible natural cutoffs as the minimal length, minimal momentum
and maximal momentum as bounds on measurement of position and
momentum. This does not mean that a particle has these features
simultaneously, but it means that measurement of position and
momentum for a test particle is bounded to these natural cutoffs.
From the mathematical grounds, the relation (\ref{eq1xp}) gives
these natural cutoffs simultaneously.
%***************************

The expression
\begin{equation}\label{dlh}
h=\frac{H}{\hbar\omega}=\frac{1}{2}(P^{2}+X^{2}),
\end{equation}
is a harmonic oscillator Hamiltonian in terms of the dimensionless
parameters with the following eigenvalue problem
\begin{equation}
h|\psi_{n}\rangle=e_{n}|\psi_{n}\rangle,      \hskip 1cm     n=0,1,2,...
\end{equation}
where $e_{n}=E_{n}/\hbar\omega$. In order to obtain the energy
eigenvalues $e_{n}$, first we show that the Hamiltonian is
factorizable, then we prove that this factorized Hamiltonian
satisfies the shape invariance condition. Now, we try to factorize
the Hamiltonian as
\begin{equation}\label{fh}
h_{0}=B^{+}(g_{0},s_{0},\nu_{0})B^{-}(g_{0},s_{0},\nu_{0})+\epsilon_{0},
\end{equation}
where
\begin{equation}
B^{\pm}(g_{0},s_{0},\nu_{0})=\frac{1}{\sqrt{2}}\Big(s_{0}X\mp
ig_{0}P\mp i\nu_{0}\Big),
\end{equation}
and $\epsilon_{0}$ is the factorization energy. Note that a new
parameter namely $\nu_{0}$, is introduced in the definition of
$B^{\pm}$ due to the existence of an additional momentum term in the
modified commutation relation. By inserting $B^{\pm}$ into the
factorized Hamiltonian (\ref{fh}), we obtain the expression
\begin{equation}
h_{0}=\frac{1}{2}[(g_{0}^{2}-2\alpha^{2} s_{0} g_{0})P^{2}+(s_{0}^{2}-\rho s_{0}g_{0})X^{2}
+(\alpha s_{0}g_{0}+2\nu_{0} g_{0})P-s_{0}g_{0}+\nu_{0}^{2}]+\epsilon_{0}.
\end{equation}
So Eq.~(\ref{dlh}) implies the following four conditions:
\begin{eqnarray}\label{c1}
g_{0}^{2}-2\alpha^{2} s_{0} g_{0}&=&1,\\
s_{0}^{2}-\rho s_{0}g_{0}&=&1,\label{c2}\\
\alpha s_{0}g_{0}+2\nu_{0}g_{0}&=&0,\label{c3}\\
\frac{1}{2}g_{0}s_{0}-\frac{1}{2}\nu_{0}^{2}&=&\epsilon_{0}.\label{c4}
\end{eqnarray}
The solutions of the above equations are given by
\begin{equation}\label{gsnu}
g_{0}=s_{0}k,  \hskip 1cm s_{0}=\frac{1}{\sqrt{1-\rho k}},\hskip 1cm
\nu_{0}=-\frac{\alpha}{2}s_{0},
\end{equation}
where
\begin{equation}\label{k}
k\equiv\frac{1}{2}(2\alpha^{2}-\rho)+\sqrt{1+\frac{1}{4}(2\alpha^{2}-\rho)^{2}}.
\end{equation}
Therefore, we can write the dimensionless Hamiltonian $h$ in the
form of Eq.~(\ref{fh}) with the factorization energy that is
obtained in Eq.~(\ref{c4}). We write a hierarchy of Hamiltonians as
\begin{equation}\label{hh}
h_{i}=B^{+}(g_{i},s_{i},\nu_{i})B^{-}(g_{i},s_{i},\nu_{i})+\sum_{j=0}^i \epsilon_{j},
\end{equation}
where $i=0,1,2,...$. In order to obtain the parameters
$s_{i},\,g_{i}$ and $\nu_{i}$, the shape invariance condition should
be satisfied
\begin{equation}\label{shc}
B^{-}(g_{i},s_{i},\nu_{i})B^{+}(g_{i},s_{i},\nu_{i})=B^{+}(g_{i+1},s_{i+1},
\nu_{i+1})B^{-}(g_{i+1},s_{i+1},\nu_{i+1})+\epsilon_{i+1}.
\end{equation}
After inserting  $B^{\pm}$ in the above equation and using the
commutation relation, we have
\begin{eqnarray}\nonumber
\frac{1}{2}[(g_{i}^{2}+2\alpha^{2} s_{i}g_{i})P^{2}+
(s_{i}^{2}+\rho  s_{i}g_{i})X^{2}
+(2\nu_{i}g_{i}-\alpha s_{i}g_{i})P+s_{i}g_{i}+\nu_{i}^{2}]=
\frac{1}{2}[(g_{i+1}^{2}-2\alpha^{2} s_{i+1}g_{i+1})P^{2}\\
+(s_{i+1}^{2}-\rho s_{i+1}g_{i+1})X^{2}
+(2\nu_{i+1}g_{i+1}+\alpha s_{i+1}g_{i+1})P-s_{i+1}g_{i+1}+\nu_{i+1}^{2}]+\epsilon_{i+1}.
\end{eqnarray}
So we find four additional conditions as:
\begin{eqnarray}\label{shc1}
g_{i+1}^{2}-2\alpha^{2} s_{i+1}g_{i+1}&=&g_{i}^{2}+2\alpha^{2}
s_{i}g_{i},\\
s_{i+1}^{2}-\rho s_{i+1}g_{i+1}&=&s_{i}^{2}+\rho  s_{i}g_{i},\label{shc2}\\
2\nu_{i+1}g_{i+1}+\alpha s_{i+1}g_{i+1}&=&2\nu_{i}g_{i}-\alpha s_{i}
g_{i},\label{shc3}\\
\frac{1}{2}(s_{i}g_{i}+s_{i+1}g_{i+1})+\frac{1}{2}(\nu_{i}^{2}-\nu_{i+1}^{2})&=&\epsilon_{i+1}.\label{shc4}
\end{eqnarray}
If we multiply Eqs.~(\ref{shc1}) and (\ref{shc2}) respectively  by
$\rho$ and $2\alpha^{2}$, we obtain
\begin{equation}
g_{i+1}^{2}-\gamma^{2} s_{i+1}^{2}=g_{i}^{2}-\gamma^{2} s_{i}^{2},
\end{equation}
where $\gamma\equiv\sqrt{\frac{2\alpha^{2}}{\rho}}$. Now it is
useful to combine the parameters $g_{i}$ and $s_{i}$ and introduce
new parameters
\begin{equation}
u_{i}=g_{i}+\gamma s_{i},   \hskip 1cm  v_{i}=g_{i}-\gamma s_{i},
\end{equation}
Their inverse transformations are
\begin{equation}\label{gs}
g_{i}=\frac{1}{2}(u_{i}+v_{i}),   \hskip 1cm
s_{i}=\frac{1}{2\gamma}(u_{i}-v_{i}),
\end{equation}
and the assumption $g_{i}, s_{i} > 0$ implies $u_{i} > |v_{i}|$.
Thus, using Eqs.~(\ref{shc1}) and (\ref{shc2}), we obtain
\begin{eqnarray}\label{uv1}
u_{i+1}^{2}+q v_{i+1}^{2}&=&v_{i}^{2}+q u_{i}^{2},\\
u_{i+1}v_{i+1}&=&u_{i}v_{i},\label{uv2}
\end{eqnarray}
where
\begin{equation}
q=\frac{1+\sqrt{2\alpha^{2}\rho}}{1-\sqrt{2\alpha^{2}\rho}}.
\end{equation}
Based on Eqs.~(\ref{uv1}) and (\ref{uv2}) let us introduce new
parameters $f_{i}$ and $t_{i}$ as
\begin{equation}
f_{i}=u_{i}v_{i},\hskip 1cm t_{i}=\frac{v_{i}}{u_{i}},
\end{equation}
where they have the same sign as $v_{i}$, and $|t_{i}|<1.$ Also we
have
\begin{equation}\label{ft1}
f_{i}=f_{0}, \hskip 1cm t_{i}=q^{-i}t_{0}, \hskip 1cm
u_{i}=q^{i/2}u_{0},  \hskip 1cm v_{i}=q^{-i/2}v_{0}.
\end{equation}
Using Eq.~(\ref{shc3}) and  $\nu_{0}$ as given in Eq.~(\ref{gsnu})
we obtain
\begin{equation}
\nu_{i+1}=\frac{1}{g_{i+1}}\left[-\alpha s_{0}g_{0}-\alpha
\left(\sum_{j=1}^{i}s_{j}g_{j}+\frac{1}{2}s_{i+1}g_{i+1}\right)\right].
\end{equation}
So we also have
\begin{equation}
\nu_{i}=\frac{-\alpha}{g_{i}}
\left(\sum_{j=0}^{i-1}s_{j}g_{j}+\frac{1}{2}s_{i}g_{i}\right).
\end{equation}
Now Eqs.~(\ref{gs}) and (\ref{ft1}) lead to
\begin{equation}
\nu_{i}=\frac{-2\alpha}{q^{i/2}u_{0}(1+\frac{t_{0}}{q^{i}})}\left\{\frac{u_{0}^{2}}{4\gamma}
\left[\left(1-\frac{t_{0}^{2}}{q^{i-1}}\right)[i]_{q}+\frac{1}{2}\left(q^{i}-\frac{t_{0}^{2}}{q^{i}}\right)\right]\right\},
\end{equation}
where we used the following definition
\begin{equation}
[i]_{q}=\frac{q^{i}-1}{q-1}.
\end{equation}
We write the eigenvalues of the Hamiltonian as
\begin{equation}
e_{n}(q,t_0)=\sum_{i=0}^n \epsilon_{i} =\sum_{i=0}^{n-1}
\epsilon_{i+1}+\epsilon_{0}
=\frac{1}{2}\sum_{i=0}^{n-1}(s_{i}g_{i}+s_{i+1}g_{i+1})
+\frac{1}{2}\sum_{i=0}^{n-1}(\nu_{i}^{2}-\nu_{i+1}^{2})+\epsilon_{0}.
\end{equation}
By writing the first term in the right-hand side of the above
equation as
\begin{equation}
\sum_{i=0}^{n-1}s_{i}g_{i}-\frac{1}{2}\sum_{i=0}^{n-1}
(s_{i}g_{i}-s_{i+1}g_{i+1}),
\end{equation}
and using the relation
$\sum_{k=0}^{n-1}(a_{k}-a_{k+1})=a_{0}-a_{n}$, we obtain
\begin{equation}
e_{n}(q,t_0)=\sum_{i=0}^{n-1}s_{i}g_{i}+\frac{1}{2}s_{n}g_{n}
-\frac{1}{2}\nu_{n}^{2}.
\end{equation}
Finally, after inserting the explicit values of the parameters, we
find the energy eigenvalues as
\begin{equation}\label{eg}
e_{n}(q,t_0)=\frac{u_{0}^{2}}{4\gamma}\left[\left(1-\frac{t_{0}^{2}}{q^{n-1}}\right)[n]_{q}+\frac{1}{2}\left(q^{n}
-\frac{t_{0}^{2}}{q^{n}}\right)\right]
\left\{1-\frac{2\alpha^{2}}{q^{n}u_{0}^{2}(1+\frac{t_{0}}{q^{n}})^{2}}
\frac{u_{0}^{2}}{4\gamma}\left[\left(1-\frac{t_{0}^{2}}{q^{n-1}}\right)[n]_{q}+\frac{1}{2}\left(q^{n}
-\frac{t_{0}^{2}}{q^{n}}\right)\right]\right\}.
\end{equation}
The first term in the above equation agrees with the harmonic
oscillator energy spectrum in the presence of the minimal
uncertainties in both position and momentum measurements \cite{21}.
Moreover, the shift in the energy spectrum due to the minimal
length, minimal momentum, and maximal momentum is smaller than the
shift in the presence of minimal length and minimal momentum.

\subsection{Special Case (limit $\rho\rightarrow0$)}
Now we consider a special case where one of the GUP parameters,
namely $\rho$ which corresponds to the minimal momentum tends to
zero. In this case, we only have the minimal length and maximal
momentum uncertainties \cite{18}
\begin{eqnarray}
[X,P]=i(1+2\alpha^{2}P^{2}-\alpha P).
\end{eqnarray}
Also, at this limit, $q$ takes the form
\begin{equation}
q\simeq1+2\sqrt{2\alpha^{2}\rho}+O(\rho),
\end{equation}
and
\begin{equation}
q^{n}\simeq1+2n\sqrt{2\alpha^{2}\rho}+O(\rho),\hskip 1cm [n]_{q}\simeq n+O(\sqrt{\rho}).
\end{equation}
The terms in the brackets in the energy spectrum (\ref{eg}) become
\begin{eqnarray}\label{spc1}
\frac{1}{4\gamma}\left(u_{0}^{2}-\frac{v_{0}^{2}}{q^{n-1}}\right)&\simeq&
g_0s_0+\frac{1}{2} (2\alpha^{2})s_0^{2}(n-1)+O(\sqrt{\rho}),\\
\label{spc2}
\frac{1}{8\gamma}\left(u_{0}^{2}q^{n}-\frac{v_{0}^{2}}{q^{n}}\right)
&\simeq&
\frac{1}{2}g_0s_0+\frac{1}{2}(2\alpha^{2})s_0^{2}n+O(\sqrt{\rho}).
\end{eqnarray}
Moreover, the parameters $g_0$ and $s_0$ are given by
\begin{equation}\label{spc3}
g_0\simeq \frac{1}{2}(2\alpha^{2})+\sqrt{1+\frac{1}{4}
(2\alpha^{2})^{2}}+O(\rho),\hskip 1 cm s_0\simeq 1+O(\rho),
\end{equation}
and
\begin{equation}
\frac{2\alpha^{2}}{q^{n}\left(u_{0}+\frac{v_{0}}{q^{n}}\right)^{2}}
\simeq
\frac{2\alpha^{2}\left(1-2n\sqrt{2\alpha^{2}\rho}\right)}{\left[2g_0(1-n\sqrt{2\alpha^{2}\rho})+2ns_0(2\alpha^{2})\right]^{2}}.
\end{equation}
Using these results, the energy spectrum takes the form
\begin{eqnarray}\nonumber\label{en}
e_{n}&\simeq&\left[\left(n+\frac{1}{2}\right)\sqrt{1+\frac{1}{4}(2\alpha^{2})^{2}}+\frac{1}{2}(2\alpha^{2})\left(n^{2}+n+\frac{1}{2}\right)\right]\\
&&\times\left\{1-\frac{\alpha^{2}(1-2n\sqrt{2\alpha^{2}\rho})}{2\left[\left(\alpha^{2}+\sqrt{1+
\alpha^{4}}\right)(1-n\sqrt{2\alpha^{2}\rho})+2n\alpha^{2}\right]^{2}}
\left[\left(n+\frac{1}{2}\right)\sqrt{1+\frac{1}{4}(2\alpha^{2})^{2}}+\frac{1}{2}(2\alpha^{2})\left(n^{2}+n+\frac{1}{2}\right)\right]\right\},\hspace{0.75
cm}
\end{eqnarray}
where the first bracket is the energy spectrum of the harmonic
oscillator in the presence of the minimal length
($e_{n}^{\mathrm{min}}$) if we replace $2\alpha^2$ with $\beta$
where $[X,P]=1+\beta P^2$ \cite{21}. Notice that, the above equation
can be written as $e_{n}=e_{n}^{\mathrm{min}}-\Delta$ where $\Delta$
is a positive term. Indeed, as it can be seen from figure 1, the
presence of the minimal length and maximal momentum decreases the
energy spectrum with respect to the existence of just the minimal
length. This effect is also observed in the perturbative study of
the problem \cite{29}.

\begin{figure}[htp]
\begin{center}
\includegraphics{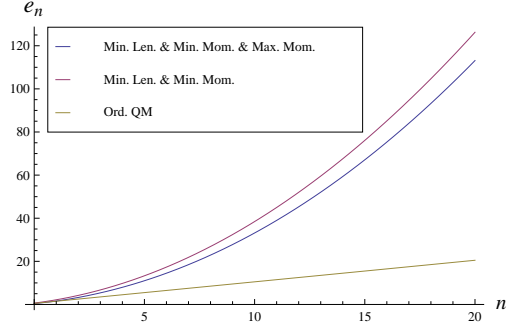}\vspace{5cm} \caption{\scriptsize{Comparing $e_{n}$ for
$\rho=10^{-4}$ and $\alpha=1$ in three scenarios.}}
\end{center}
\end{figure}

\subsection{Supersymmetric partner}
Using Eq.~(\ref{hh}) the Hamiltonians  of the SUSYQM hierarchy can
be written as
\begin{equation}
h_{i}=\frac{1}{2}\left[(g_{i}^{2}-2\alpha^{2}g_{i}s_{i})P^{2}+(s_{i}^{2}-\rho g_{i}s_{i})X^{2}+(2g_{i}\nu_{i}
+\alpha g_{i}s_{i})P-g_{i}s_{i}+\nu_{i}^{2}\right]+\sum_{j=0}^i\epsilon_{j}.
\end{equation}
It can also be expressed as
\begin{equation}
h_{i}=\frac{1}{2}(a_{i}P^{2}+b_{i} X^{2})+c_{i}P+d_{i},
\end{equation}
where
\begin{eqnarray}
a_{i}&=&g_{i}^{2}-2\alpha^{2}g_{i}s_{i}
=\frac{u_{0}^{2}}{2(q+1)}(q^{i}+t_{0})\left(1+\frac{t_{0}}{q^{i-1}}\right),\\
b_{i}&=&s_{i}^{2}-\rho g_{i}s_{i}=
\frac{u_{0}^{2}}{2\gamma^{2}(q+1)}(q^{i}-t_{0})\left(1-\frac{t_{0}}{q^{i-1}}\right),\\
c_{i}&=&2g_{i}\nu_{i}+\alpha g_{i}s_{i}=-u_{0}^{2}\sqrt{\frac{\rho}{8}}\left(1-\frac{t_{0}^{2}}{q^{i-1}}\right)[i]_{q},\\
d_{i}&=&\sum_{j=0}^i\epsilon_{j}-\frac{1}{2}g_{i}s_{i}+\frac{1}{2}\nu_{i}^{2}
=\frac{u_{0}^{2}}{4\gamma}\left(1-\frac{t_{0}^{2}}{q^{i-1}}\right)[i]_{q}.
\end{eqnarray}
So the Hamiltonian in terms of the variables with dimensions are
given by
\begin{equation}\label{Hi}
H_{i}\equiv\hbar\omega h_{i}=\frac{p^{2}}{2m_{i}}+
\frac{1}{2}m_{i}\omega_{i}^{2}x^{2}+
c_{i}\sqrt{\frac{\hbar\omega}{m}}p+d_{i}\hbar \omega,
\end{equation}
where $m_{i}=m/a_{i}$ and $\omega_{i}=\sqrt{a_{i}b_{i}}\omega$.
Equivalently we have
\begin{equation}
H_{i}=\frac{1}{2m_{i}}\left(p+c_{i}\sqrt{\frac{m_{i}\hbar\omega}{a_{i}}}\right)^{2}+
\frac{1}{2}m_{i}\omega_{i}^{2}x^{2}-\frac{1}{2}\left(\frac{c_{i}^{2}}{a_{i}}-2d_{i}\right)\hbar\omega,
\end{equation}
Therefore, in the GUP framework, the harmonic oscillator belongs to
a hierarchy of Hamiltonians with a shift in momentum space and with
different masses and frequencies.

\section{One-dimensional harmonic oscillator in a uniform electric field}
Now let us study a particle with mass $m$ and charge $\bar{q}$ in
the presence of the harmonic oscillator potential
$V(x)=\frac{1}{2}m\omega^{2}x^{2}$ and a uniform electric field
$\bar{\varepsilon}$. The  Hamiltonian of this system is
\begin{equation}
H=\frac{p^{2}}{2m}+\frac{1}{2}m\omega^{2}x^{2}-\bar{q}\bar{\varepsilon}x,
\end{equation}
which can be written in terms of the dimensionless operators as
\begin{equation}\label{ufh}
h=\frac{H}{\hbar\omega}=\frac{1}{2}(P^{2}+X^{2})-\varepsilon X,
\end{equation}
where $\varepsilon=\bar{q}\bar{\varepsilon}a/(\hbar\omega)$. Again,
in order to find the exact energy spectrum we factorize the
Hamiltonian as
\begin{equation}\label{hh1}
h_{0}=B^{+}(g_{0},s_{0},\nu_{0},r_{0})B^{-}(g_{0},s_{0},\nu_{0},r_{0})+\epsilon_{0},
\end{equation}
where $\epsilon_{0}$, is the factorization energy and $B^{\pm}$
are defined as
\begin{equation}\label{B+}
B^{\pm}(g_{0},s_{0},\nu_{0},r_{0})=\frac{1}{\sqrt{2}}(s_{0}X\mp ig_{0}P\mp i\nu_{0}+r_{0}).
\end{equation}
Here $r_0$ is the new parameter that corresponds to the nonzero
electric field. Now we have
\begin{equation}
h_{0}=\frac{1}{2}[(g_{0}^{2}-2\alpha^{2} g_{0}s_{0})P^{2}+(s_{0}^{2}-\rho g_{0}s_{0})X^{2}+(2g_{0}\nu _{0}+\alpha s_{0}g_{0})P+
2r_{0} s_{0}X-g_{0}s_{0}+r_{0}^{2}+\nu_{0}^{2}]+\epsilon_{0}.
\end{equation}
This equation yields five conditions on the parameters which three
of them are the same as Eqs.~(\ref{c1}-\ref{c3}). The remaining
conditions are
\begin{eqnarray}\label{rs1}
r_{0}s_{0}&=&-\varepsilon,\\
\frac{1}{2}(g_{0}s_{0}-r_{0}^{2}-\nu_{0}^{2})&=&\epsilon_{0}.
\end{eqnarray}
Also the hierarchy of Hamiltonians is
\begin{equation}
h_{i}=B^{+}(g_{i},s_{i},\nu_{i},r_{i})B^{-}(g_{i},s_{i},\nu_{i},r_{i})+\sum_{j=0}^i
\epsilon_{j},
\end{equation}
and the shape invariance condition reads
\begin{equation}
B^{-}(g_{i},s_{i},\nu_{i},r_{i})B^{+}(g_{i},s_{i},\nu_{i},r_{i})
=B^{+}(g_{i+1},s_{i+1},\nu_{i+1},r_{i+1})B^{-}(g_{i+1},s_{i+1},\nu_{i+1},r_{i+1})+\epsilon_{i+1},
\end{equation}
Similar to the absence of the electric field, the above equation
gives five conditions on the parameters which three of them are the
same as Eqs.~(\ref{shc1}-\ref{shc3}) and the other two conditions
are given by
\begin{eqnarray}\label{rs2}
r_{i+1}s_{i+1}&=&r_{i}s_{i},\\
\epsilon_{i+1}&=&\frac{1}{2}(s_{i}g_{i}+s_{i+1}g_{i+1})+\frac{1}{2}
(\nu_{i}^{2}-\nu_{i+1}^{2})+\frac{1}{2}(r_{i}^{2}-r_{i+1}^{2}).\label{egc}
\end{eqnarray}
Now following Ref.~\cite{27} parameters $r_{i}$ read
\begin{equation}
r_{i}=\frac{-2\gamma\varepsilon}{u_{0}}q^{-i/2}\left(1-\frac{t_{0}}{q^{i}}\right)^{-1}.
\end{equation}
In addition, Eq.~(\ref{egc}) gives  the energy spectrum as
\begin{equation}\label{egr}
e_{n}(q,t_{0},\varepsilon)=\sum_{i=0}^n \epsilon_{i}=\sum_{i=0}^{n-1}s_{i}g_{i}+\frac{1}{2}s_{n}g_{n}
-\frac{1}{2}\nu_n^2
-\frac{1}{2}r_n^2=e_{n}(q,t_{0},0)+\Delta e_{n}(q,t_{0},\varepsilon),
\end{equation}
where $e_{n}(q,t_{0},0)$ is the energy spectrum in the absence of
the electric field and $\Delta e_{n}(q,t_{0},\varepsilon)$, is the
shift due to the presence of the electric field
\begin{equation}
\Delta e_{n}(q,t_{0},\varepsilon)=-\frac{2\gamma^{2}\varepsilon^{2}}{u_{0}^{2}}q^{-n}\left(1-\frac{t_{0}}{q^{n}}\right)^{-2}.
\end{equation}
Finally, the Hamiltonians of the SUSYQM hierarchy can be written as
\begin{equation}
H_{i}\equiv\hbar\omega h_{i}=\frac{1}{2m_{i}}\left(p+c_{i}\sqrt{\frac{m_{i}\hbar\omega}{a_{i}}}\right)^{2}+
\frac{1}{2}m_{i}\omega_{i}^{2}x^{2}-\frac{1}{2}\left(\frac{c_{i}^{2}}{a_{i}}-2d_{i}\right)\hbar\omega
-\bar{q}\bar{\varepsilon}x.
\end{equation}
Thus, in the presence of a uniform electric field, the harmonic
oscillator belongs to a hierarchy of Hamiltonians with a shift in
momentum space and with different masses and frequencies but with
the same electric field.

\section{Conclusions}
In this Letter, we have considered a GUP framework that admits
maximal momentum uncertainty and nonzero minimal position and
momentum uncertainties. We applied the supersymmetric quantum
mechanics method and the shape invariance condition in order to find
the exact GUP-corrected harmonic oscillator energy spectrum without
solving the corresponding generalized Schr\"odinger equation. The
results show that although the shift in the energy spectrum is
positive, it is smaller with respect to the case with minimal
position and momentum uncertainties. We obtained the supersymmetric
partner Hamiltonians and showed that the GUP-corrected harmonic
oscillator belongs to a hierarchy of Hamiltonians of the same type
but with a shift in momentum space and with different masses and
frequencies. Finally, we have studied the effects of a uniform
electric field on this modified harmonic oscillator energy spectrum.

\section*{Acknowledgements}

The work of  K. Nozari has been supported financially by the Center
for Excellence in Astronomy and Astrophysics (CEAAI - RIAAM),
Maragha, Iran.

\end{document}